\documentclass[12pt]{article}

\usepackage{amsmath, amssymb,amsthm,amscd,graphicx,dsfont}
\usepackage{hyperref,subfigure}
\usepackage{bbding,phaistos,wasysym,bbm,mathscinet}
\usepackage[mathscr]{eucal}

\newcommand{\address}[1]{\vbox{\center\em#1}}

\numberwithin{equation}{section}
\theoremstyle{plain} 
\theoremstyle{definition}
\newtheorem{thm}{Theorem}[section]

\theoremstyle{definition}
\newtheorem{defn}{Definition}[section]

\newtheorem{conj}{Conjecture}[section]

\theoremstyle{remark}

\newcommand{\R}{\mathbb{R}}
\newcommand{\C}{\mathbb{C}}

\renewcommand{\O}{\mathcal{O}}

\newcommand{\bra}[1]{\langle #1|}
\newcommand{\ket}[1]{|#1\rangle}
\newcommand{\braket}[2]{\langle #1|#2\rangle}
\newcommand{\braOket}[3]{\langle #1| #2 | #3\rangle}

\newcommand{\pp}[2]{\frac{\partial #1}{\partial #2}}

\DeclareMathOperator{\Tr}{Tr}
\DeclareMathOperator{\Spec}{Spec}
 
\newcommand*{\BigBox}{\fboxsep 0pt\fbox{\rule{0.1em}{1.5 em}\rule{1.5 em}{0.1em}}}
\newcommand*{\BBox}{ \raisebox{-5pt}{\BigBox}}
\def\sector#1#2{ \raise 1pt \hbox{$\scriptstyle #1$}  \hskip 0mm
 \mathop{\BBox}\limits_{\raise 6pt \hbox{$\scriptstyle#2$}}\hskip 1mm}

\DeclareMathOperator{\End}{\mathrm{End}}
\DeclareMathOperator{\Hom}{\mathrm{Hom}}

\DeclareMathOperator{\Mod}{\mathrm{mod}}

\newcommand{\Coh}{\mathrm{Coh}} 

\usepackage{mathrsfs}

\DeclareMathOperator{\Vol}{\mathrm{Vol}}

\begin{document}
\begin{titlepage}

\begin{flushright}
IPMU 10-0198
\end{flushright}

\begin{center}
\vskip 10mm
{
\Large
Equivalence of A-Maximization and Volume Minimization}
\vskip 10mm
Richard Eager${}$\footnote{\href{mailto:reager@physics.ucsb.edu}{\rm
reager@physics.ucsb.edu}}
\vskip 5mm
\address{
${}$
Department of Physics \\
University of California \\
Santa Barbara, CA 93106, USA \\
and \\
Institute for the Physics and Mathematics of the Universe \\
The University of Tokyo \\
Kashiwa, Chiba 277-8582, Japan
}
\vskip 8mm
\end{center}

\abstract{
The low energy effective theory on a stack of D3-branes at a Calabi-Yau singularity is an $\mathcal{N} = 1$ quiver gauge theory.  The AdS/CFT correspondence predicts that the strong coupling dynamics of the gauge theory is described by weakly coupled type IIB supergravity on $AdS_5 \times L^5,$ where $L^5$ is a Sasaki-Einstein manifold.  Recent results on Calabi-Yau algebras efficiently determine the Hilbert series of any superconformal quiver gauge theory.  We use the Hilbert series to determine the volume of the horizon manifold in terms of the fields of the quiver gauge theory.  One corollary of the AdS/CFT conjecture is that the volume of the horizon manifold $L^5$ is inversely proportional to the a-central charge of the gauge theory.   By direct comparison of the volume determined from the Hilbert series and the a-central charge, this prediction is proved independently of the AdS/CFT conjecture.
\noindent \normalsize{
 } }

\vfill

\end{titlepage}

\setcounter{footnote}{0}

{\addtolength{\parskip}{-1ex}\tableofcontents}

\newpage
\section{Introduction}
Maldacena's original AdS/CFT correspondence relates type IIB string theory on $AdS_5 \times S^5$ to $\mathcal{N} = 4$ supersymmetric Yang-Mills in Minkowski space.  Several authors \cite{Morrison:1998cs,Acharya:1998db} realized that this correspondence could be extended to cases with fewer supersymmetries.  If the five-sphere is replaced by another five-dimensional manifold $L^5,$ $\mathcal{N} = 1$ SUSY is preserved only if $L^5$ is Sasaki-Einstein.  For these manifolds Gubser, \cite{Gubser:1998vd, Henningson:1998gx} proposed a simple yet powerful prediction of the AdS/CFT correspondence.  Proper normalization of the AdS 3-point functions ensures that the volume of the Sasaki-Einstein manifold is inversely proportional to the central charge $a,$
$$\Vol(L^5) \propto \frac{\pi^3}{4} \frac{1}{a}.$$

The a-central charge of a 4D SCFT quiver gauge theory can be determined through a variational procedure called a-maximization developed by Intriligator and Wecht \cite{Intriligator:2003jj}.  Martelli, Sparks, and Yau \cite{Martelli:2005tp,Martelli:2006yb} proposed that the dual variational problem is minimizing the volume of the horizon manifold over all possible choices of a ``Reeb'' vector.

We show the equivalence of these two procedures by describing volume minimization in terms of the fields of the quiver.  The volume of the horizon manifold is governed by the asymptotic growth of the number of holomorphic functions on its metric cone $X = C(L^5)$ \cite{Bergman:2001qi}.  Using the correspondence between holomorphic functions on $X$ and mesonic operators in the quiver, we express the Hilbert series in terms of mesonic operators.  Using this correspondence, we formulate volume minimization entirely in terms of the fields of the quiver gauge theory.  We will perturbatively expand the expression for the volume.  Several terms in the expression vanish from constraints from $\mathcal{N} = 1$ superconformal field theories.  After the cancellations are accounted for, we will see that the expressions for volume minimization and a-maximization are identical.

For toric Calabi-Yau singularities, the relationship between the $a$-central charge and volume has already been established
\cite{Butti:2005ps, Lee:2006ru}.  Our proof applies to both toric and non-toric singularities.  While branes at toric singularities have been extensively studied \cite{Franco:2005rj}, far less is known about branes at general Calabi-Yau singularities.

Our plan for the paper is as follows.
We first review the relation between the volume and a-central charge predicted by the AdS/CFT correspondence.  Next we introduce the general structure of quiver gauge theories and explain the role of baryonic symmetries in quiver gauge theories and their supergravity duals.  The subsequent sections form the mathematical core of this paper.  Section \ref{sec:CY} introduces Calabi-Yau algebras, which mathematically characterize quiver gauge theories that flow to $\mathcal{N} = 1$ superconformal field theories in the infrared.  The next section introduces the stronger notion of a ``non-commutative crepant resolution.''  Non-commutative crepant resolutions describe the $\mathcal{N} = 1$ superconformal field theories which can be engineered from D3-branes at Calabi-Yau singularities.  These will be the main source of Calabi-Yau algebras in this paper.  Using the projective resolution of modules, a property satisfied by Calabi-Yau algebras, we will explain how to to compute the Hilbert series of a quiver gauge theory in section \ref{sec:Hilbert}.  Examples of Hilbert series are given in section \ref{sec:examples}.  We review the gauge theories associated to $\C^3$ and the conifold and show how the Hilbert series correctly determines the volume of their horizon manifolds.  Finally in section \ref{sec:pert}, we prove the equivalence of a-maximization and volume minimization for general quiver gauge theories.
\section{Predictions from AdS/CFT}
The AdS/CFT correspondence between type IIB string theory with $N$ D3-branes at a local Calabi-Yau singularity $X$ and type IIB string theory on $AdS_5 \times L^5$ leads to a rich interplay between gauge theory, supergravity, and mathematics.  In the low-energy limit, the correspondence is a realization of holography \cite{tHooft:1993gx, Susskind:1994vu}.  At low energies, the correspondence is between a gauged supergravity theory on $AdS_5$ and a superconformal field theory living on the boundary of $AdS_5.$  We focus on the limit where the number, $N,$ of D3-branes is large.  For the low energy effective field theory on the D3-brane world-volume to have $\mathcal{N} = 1$ supersymmetry, $X$ must be Calabi-Yau, possibly with Gorenstein singularities.  We will consider only isolated Gorenstein\footnotemark \;
singularities so that the near horizon limit can easily be defined.
\footnotetext{An isolated complex threefold singularity is {\it Gorensein} if it has a no-where vanishing holomorphic three form $\Omega^{3,0}$ that is well-defined away from the singular point.} 
Furthermore, we only consider Gorenstein singularities that can be realized as a metric cone over a Sasaki-Einstein base $L^5.$  As emphasized in \cite{Gauntlett:2006vf}, not all Gorenstein singularities satisfy this property.  For the supergravity theory to have $\mathcal{N} = 1$ supersymmetry, $L^5$ must be Sasaki-Einstein.  An odd dimensional Riemannian manifold $L$ is Sasakian if its metric cone $(C(L), g_L)$ with
$$g_{C(L)} = dr^2 + r^2 g_L$$
is K\"ahler.  The K\"ahler condition implies that $C(L)$ contains an almost-complex structure $J.$  If additionally the metric cone $C(L)$ is a possibly singular Calabi-Yau, then $L$ is called Sasaki-Einstein.  Every Sasaki-Einstein manifold posses a distinguished vector field
$$\xi = J \left(r \pp{}{r}  \right)$$ called the Reeb vector field.  The symmetry generated by the Reeb vector field is dual to the $R$-symmetry of the superconformal gauge theory.  If the orbits of the Reeb vector close, then $L^5$ is either regular or quasi-regular.  This is dual to the field theory having a compact $R$-symmetry group, isomorphic to $U(1)$ .  If the orbits of the Reeb vector do not close, then $L^5$ is an irregular Sasaki-Einstein manifold and the $R$-symmetry group of the dual gauge theory is non-compact and isomorphic to $\R$.

The AdS/CFT correspondence matches the isometries of the supergravity theory to global symmetries of the dual four dimensional superconformal field theory.  The four dimensional superconformal algebra $psu(2,2|1)$ contains the bosonic subalgebra $so(4,2) \times u(1)_R.$  Under the AdS/CFT correspondence, the $SO(4,2)$ global symmetry group matches the isometry group of $AdS_5.$  Every Sasaki-Einstein manifold has a symmetry generated by the Reeb vector field.  Under the AdS/CFT correspondence, this symmetry maps to the $R$-symmetry of the field theory.  We will consider the dimensional reduction of IIB supergravity on $L^5.$  There are $b^3(L^5)$ gauge fields $A^I, I = 1, \dots b^3(L^5)$ from dimensional reduction of the RR four-form.  There is an additional $U(1)$ gauge field from the Kaluza-Klein reduction of the graviton.  If $L^5$ possesses isometries in addition to the one generated by the Reeb vector field, then the field theory has additional mesonic flavor symmetries \cite{Barnes:2005bw, Benvenuti:2006xg}, which we will review in section \ref{sec:baryon}.  Under the AdS/CFT correspondence, the bulk gauge fields correspond to global symmetries of the boundary field theory.  In addition to the matching of symmetries, the AdS/CFT correspondence predicts a precise relationship between correlation functions.

Suppose the $AdS_5$ theory has gauge group $G$ of rank $|G|$ and gauge fields $A^{I}, I = 1, \dots |G|.$  The gauge symmetries are mapped to global symmetries of the boundary theory with corresponding currents $J_I.$  Gubser, Klebanov, Polyakov, and Witten \cite{Gubser:1998bc, Witten:1998qj} proposed the following way to match partition functions between the CFT and SUGRA theories. Background gauge fields $A^I_{0}$ turned on in the CFT can be extended to gauge fields $A^I$ in the interior of  $AdS_5$ in a unique manner up to gauge transformations.  The partition function of the CFT with background fields $A^I_{0}$ equals the SUGRA partition function with the restriction that the components of the dynamical gauge fields $A^I$ approach the CFT background fields $A^I_0$ at the boundary of $AdS_5.$  We schematically represent this as 
$$ Z[A^{I}_0]_{CFT} = Z_{SUGRA}[A^I \vert_{\partial AdS_5} = A^I_0].$$
Here the CFT generating functional is
$$ Z[A^I_0]_{CFT} = \Big \langle \exp  \left( \int J_I A^I_0 \right) \Big \rangle_{CFT}.$$

Under the GKP/W prescription, the gauge symmetry of the $AdS$ gauge fields, $A_I \rightarrow A_I + \partial \chi_{I},$ translates directly into the condition that the CFT currents are conserved, $\partial_{\mu} J^{\mu}_I = 0.$  Since the AdS/CFT correspondence is a weak-strong duality, it is usually difficult to test the equivalence of correlation functions.  For the original AdS/CFT correspondence with $\mathcal{N} = 4$ supersymmetry, the additional supersymmetry has enabled extensive tests of the correspondence.  For theories with only $\mathcal{N} = 1$ supersymmetry, there are very few quantities we can compute at strong coupling.  However, we can still try to match global anomalies, which are one-loop exact and therefore computable at strong coupling.  The $U(1)$ global symmetries are exact symmetries of the quantum theory.  When coupled to external gauge fields, these symmetries can have ABJ \cite{Adler:1969gk, Bell:1969ts} type triangle anomalies.

A direct check of AdS/CFT can be made by showing that the the three-point functions on both sides of the correspondence match.  For anomalies, there is an elegant method that is equivalent to matching the three-point functions of the anomalous currents.  Witten \cite{Witten:1998qj} observed that the 5d Chern-Simons term in the $AdS_5$ supergravity action is not gauge invariant.  Under a gauge transformation, the 5d Chern-Simons term gains a boundary term.  Under the GKP/W prescription, this term becomes precisely the 4D ABJ anomaly in the boundary SCFT.  

Four dimensional superconformal field theories are parametrized by two central charges, $a$ and $c$.  The central charges can be read off from the two- and three-point function of the stress energy tensor.  Alternatively, the anomaly coefficients can be computed from Weyl anomalies.  Since the stress energy tensor is a composite operator, it must be appropriately regularized.  Conformal symmetry requires that the trace of the stress tensor vanishes.  However, the trace and regularization procedures do not commute, and their failure to do so leads to the Weyl anomaly.  For any theory with a large $N$ holographic dual, the $a$ and $c$ central charges must be equal \cite{Henningson:1998gx}.  This is automatically the case for superconformal quiver gauge theories \cite{Intriligator:2003wr} \cite{Benvenuti:2004dw}.  The difference $a - c$ is proportional to $\Tr R = 0$ to leading order in $N.$  For a superconformal quiver, the condition $\Tr R = 0$ can be seen by taking the linear combination of the NSVZ beta functions \cite{Novikov:1983ee} weighted by the ranks of the gauge groups.

Since the stress energy tensor and the $R$-symmetry current both reside in the same supersymmetry multiplet, the $a$ central charge can be written as
$$a = \frac{3}{32} \left( 3 \Tr R^3 - \Tr R \right).$$
The trace is over all the fields, and $R$ is the $R$-charge under the IR $R$-symmetry. 

Either by matching 3-point functions or generalizing Witten's argument, the AdS/CFT correspondence predicts that the volume of the Sasaki-Einstein manifold is inversely proportional to the central charge $a,$
$$\Vol(L^5) = \frac{\pi^3 N^2}{4 a}.$$
After reviewing the general properties of quiver gauge theories, we will explain how the a-central charge is determined by Intriligator and Wecht's a-maximization procedure.
\section{Quiver Gauge Theories}
The world-volume gauge theory on a stack of D3-branes at a Calabi-Yau singularity is often described by a quiver gauge theory.  A {\it quiver} $Q = (V,A,h,t: A \rightarrow V)$ is a collection of vertices $V$ and arrows $A$ between the vertices of the quiver.  The arrows are directed edges with the head and tail of an arrow $a \in A$ given by maps $h(a)$ and $t(a),$ respectively.  A {\it representation} $X$ of a quiver is an assignment of $\C$-vector spaces $X_{v}$ to every vertex $v \in V$ and a $\C-$linear map $\phi_{a}: X_{t(a)} \rightarrow X_{h(a)}$ to every arrow $a \in A.$  The {\it dimension vector} $\mathbf{n} \in \mathbb{N}^{|V|}$ of a representation $X$ is a vector with an entry for each vertex $v \in V$ equal to the dimension of the vector space $X_v.$

A {\it quiver gauge theory} is specified by a quiver and a superpotential in the following manner:
\begin{itemize}
\item The gauge group
$$G = \prod_{v \in V} U(n_v)$$
is a product of unitary groups $U(n_v)$ of dimension $n_v.$
\item Arrows $a \in A$ represent chiral superfields $\Phi_a$ transforming in the fundamental representation of $U(n_{h(a)})$ and in the anti-fundamental representation of $ U(n_{t(a)})$.  If the two vertices are distinct, the chiral superfields are called {\it bifundamental} fields.  Otherwise, the arrow is a loop and the field transforms in the adjoint representation.
\item The superpotential
$$W = \sum_{l = a_1 a_2 \dots a_k \in L} \lambda_{l} \Tr \left[ \Phi_{a_1} \Phi_{a_2} \dots \Phi_{a_k} \right]$$
is a sum of gauge invariant operators $\Tr \left[ \Phi_{a_1} \Phi_{a_2} \dots \Phi_{a_k} \right].$  Gauge invariance requires $l = a_1 a_2 \dots a_k$ to be an oriented loop in the quiver.  Each operator has coupling constant $\lambda_{l}.$
\end{itemize}

For a quiver gauge theory to be physically sensible, the gauge anomalies for each gauge group must vanish.
Vanishing of the triangle anomaly with three external gluons of the $U(n_v)$ gauge group yields the condition
\begin{equation}
\sum_{a \in A | h(a) = v} n_{t(a)} - \sum_{a \in A | t(a) = v} n_{h(a)} = 0.
\label{eqn:rank}
\end{equation}
Linear combinations $U(1)_q$ of the $U(1)_v \subset U(n_v)$ groups can mix and lead to triangle anomalies of the form $\Tr \left[ SU(n_v)^2 U(1)_q \right].$  Vanishing of this mixed anomaly requires
\begin{equation}
\label{eqn:baryon}
\sum_{a \in A | h(a) = v} n_{t(a)} q_{t(a)} - \sum_{a \in A | t(a) = v} n_{h(a)} q_{n(a)} = 0.
\end{equation}
Quiver gauge theories describing the low energy effective field theory of D-branes at a Calabi-Yau singularity have a variant of the Green-Schwarz mechanism to cancel the anomalous $U(1)$'s.
The gauge fields of the anomalous $U(1)$'s couple to RR-form fields giving them St\"{u}ckelberg masses \cite{Douglas:1996sw, Ibanez:1998qp, Antoniadis:2002cs}.  These massive vector fields decouple in the IR.  The non-anomalous $U(1)$ fields are free in the infrared so they also decouple and become global $U(1)$ symmetries in the IR.  These global $U(1)$ symmetries are called baryonic symmetries.  This is explained from a large-volume perspective in \cite{Jockers:2004yj, Buican:2006sn, Martelli:2008cm}.  In the next section we will review baryonic symmetries in more detail.

At a conformal fixed point in the infrared, we expect the NSVZ 1-loop exact beta functions of the gauge groups $SU(n_v)$ and couplings $\lambda_l$ to vanish.  These constraints are
\begin{align}
\label{NSVZ}
\hat{\beta}_{1/g_v^2} & = 0 & 2n_v + \sum_{e \in Q_1} (R(e)-1) n_{t(e)} + \sum_{e \in Q_1} (R(e) - 1) n_{h(e)} & = 0\\
\hat{\beta}_{\lambda_l} & = 0 & -2 + \sum_{e \in \text{loop }l} R(e) & = 0.
\end{align}
The last condition implies that at a superconformal fixed point, every term in the superpotential has total $R$-charge 2.
\section{Baryonic and Flavor Symmetries}
\label{sec:baryon}
Global flavor symmetries play a prominent role in our story because they can mix with the $R$-symmetry of the superconformal gauge theory.  The a-maximization procedure of Intriligator and Wecht determines the precise form of the mixing.  In this section, we review the constraints on anomalies with flavor symmetries.  These constraints will be essential when we analyze the perturbative expansion of the Hilbert series in section \ref{sec:Hilbert}.

After dimensional reduction, D3 branes wrapping 3-cycles in $L^5$ become baryonic particles in the $AdS_5$ supergravity theory.  They are charged under the $b^3(L^5)$ gauge fields coming from dimensional reduction of the RR 4-form on the same cycle.  Under the AdS/CFT correspondence, these gauge fields are dual to global baryonic $U(1)$ symmetries.  For quiver gauge theories, the baryonic symmetries can be described by charges $q_v^I$ satisfying equation \eqref{eqn:baryon}.  The charge of a bifundamental field $X_{t(a),h(a)}$ under the $I^{th}$ global baryonic symmetry is $B^I(X) = q_{h(a)}^I - q_{t(a)}^I.$
When $q_v = 1,$ none of the bifundamental fields is charged under the baryonic symmetry.  In this case, \eqref{eqn:baryon} becomes equivalent to \eqref{eqn:rank}.  The other solutions have non-vanishing baryonic charges, so the dimension of the solution space of \eqref{eqn:baryon} is $b^3(L^5) + 1.$

Mesonic operators in the quiver gauge theory are uncharged under baryonic symmetries.  However they are charged under the $R$-symmetry and possibly additional flavor symmetries.  If $L^5$ has a rank 
$\ell$-dimensional space of isometries, then there are $\ell$ Kaluza-Klein gauge fields in the $AdS_5$ supergravity theory \cite{Barnes:2005bw, Benvenuti:2006xg}.  The Kaluza-Klein gauge fields are dual to non-baryonic flavor symmetries in the SCFT.  These symmetries are called {\it mesonic flavor symmetries} because mesons are charged under them.

In addition to the anomalies \eqref{eqn:mixed}, the baryonic symmetries of four dimensional superconformal field theories satisfy relations:
\begin{align}
\Tr B^I & = 0 \\
\label{eqn:cubic}
\Tr B^I B^J B^K & =  0 \qquad  \text{for all } I,J,K.
\end{align}
since there are no 10-dimensional Chern-Simons couplings that could generate the corresponding anomalies via dimensional reduction \cite{Intriligator:2003wr, Martelli:2008cm}.
\section{A-Maximization}
\label{sec:Amax}
Given the ultraviolet description of a quiver gauge theory, determining the exact $R$-symmetry in the IR is complicated by the possibility that the $R$-symmetry can mix with other $U(1)$ global symmetries.  Intriligator and Wecht \cite{Intriligator:2003jj} developed a procedure called a-maximization to determine the true $R$ symmetry in the IR.  They first consider a trial $R$-symmetry
$$R_t = R_0 + \sum_{I} s^I F^I$$
where $R_0$ is any $U(1)$ charge assignment whose gauge and superpotential couplings have vanishing beta functions \eqref{NSVZ}.  The $F^I$ represent arbitrary $U(1)$ flavor symmetries and $s^I$ are parameters.  Combined with the general results on flavor symmetries in $\mathcal{N} = 1$ SCFTs \cite{Anselmi:1997am},
\begin{align}
\label{eqn:mixed}
9 \Tr (R^2 F^I) & = \Tr F^I \\
\Tr R F^J F^K & \text{ is negative definite.}
\end{align}
Intriligator and Wecht showed that the true $R$ symmetry is the one that minimizes the 4D central charge
$$a = \frac{3}{32} \left( \sum_{\psi} 3 R_{\psi}^3 - R_{\psi} \right).$$
Since the a-central charge can be expressed in terms of triangle anomalies, the sum is over all fermions, $\psi,$ in the quiver gauge theory.  A chiral multiplet $X_e$ containing a complex scalar field with R-charge $R(e)$ also contains a fermion with R-charge $R(e) - 1.$  Bifundamental fields between gauge groups of ranks $n_v$ and $n_w$ contribute $n_v n_w$ fermions to the gauge theory matter content.  Similarly, adjoint fields contribute $n_v^2$ fermions.
For each gauge group $U(n_v),$ there are $n_v^2$ gauginos, which all have R-charge 1.  In terms of the fields of the quiver, the a central charge is
\begin{equation*}
a = \frac{3}{32} \left( 2 N_G +  \sum_{e \in \text{Arr}(v \rightarrow w)}  3 n_v n_w (R(e)-1)^3 - n_v n_w (R(e) - 1) \right)
\end{equation*}
where $N_G = \sum_{v \in Q_0} n_v^2$ is the number of gauginos.  For a superconformal quiver gauge theory $\Tr R = 0,$ which lets us write the a-anomaly as
\begin{equation}
\label{eqn:acentral}
a = \frac{9}{32} \left( N_G +  \sum_{e \in \text{Arr}(v \rightarrow w)}   n_v n_w (R(e) - 1)^3  \right).
\end{equation}

As emphasized in \cite{Butti:2005vn, Lee:2006ru, Martelli:2008cm} the baryonic symmetries decouple from the maximization procedure, so we can restrict the parameters $s^I$ to vary over the $\ell$-dimensional subspace of mesonic flavor symmetries in a-maximization.  The space of mesonic flavor symmetries corresponds directly to the $\ell$-dimensional subspace the Reeb vector is varied over in volume minimization.
We have given an account of the original Intriligator-Wecht procedure, which is sufficient for our purposes.  For further developments and modifications, see \cite{Kutasov:2003ux,Bertolini:2004xf,Benvenuti:2004dy}.
\section{Calabi-Yau Algebras}
\label{sec:CY}
Which quiver gauge theories arise from placing a stack of D3-branes at a Calabi-Yau singularity?  Berenstein and Douglas \cite{Berenstein:2002fi} suggested that the Calabi-Yau condition should be captured by a form of Serre duality.  Additionally, they conjectured that the Calabi-Yau condition could be captured by a projective resolution of simple modules.  In this section, we will review the homological algebra necessary to state Ginzburg's version \cite{vdBtalk, ginzburgcy} of Berenstein and Douglas' conjecture.  We will be able to use Ginzburg's projective resolution to determine the Hilbert series of any Calabi-Yau algebra of dimension three.

Following \cite{broomthesis}, let $S := \bigoplus_{v \in Q_0} \C e_v$ be the semi-simple algebra generated by the paths of length zero.  Similarly, let $T_1 = \bigoplus_{a \in Q_1} \C x_a$ be the vector space generated by the arrows.  For each arrow $a \in Q_1,$ there is a relation $R_a \equiv \pp{}{x_a} W$.  Define $T_2 = \bigoplus_{a \in Q_1} \C R_a$ to be the vector space generated by the relations $R_a \equiv \pp{}{x_a} W.$  In addition to relations, there can also be relations between relations called syzygies.  For any superpotential algebra, there is a universal syzygy \cite{MR1247289} associated to every vertex $v \in Q_0$ of the form
$$W_v := \sum_{a \in Q_1 | t(a) = v} x_a R_a = \sum_{a \in Q_1 | h(a) = v} R_a x_a.$$
Finally, let $T_3 := \bigoplus_{v \in Q_0} \C W_v$ be the vector space spanned by the universal syzygies.  There are natural maps $\mu_0, \dots \mu_3$ between these spaces.
The map $\mu_0$ takes two paths and concatenates them.  It is extended by linearity to act on the entire path algebra:
\begin{align*}
\mu_0 :  A & \otimes_S A \rightarrow A \\
 x &\otimes y \rightarrow xy. \\
\end{align*}
The map $\mu_1$ is defined on a triple (path, arrow, path) and produces a formal difference of pairs of paths.  By linearity the map extends to the entire path algebra.
\begin{align*}
\mu_1 :  A & \otimes_S T \otimes_S A \rightarrow A \otimes_S A \\
 x &\otimes x_a \otimes y \rightarrow xx_a \otimes y - x \otimes x_a y. \\
\end{align*}
The map $\mu_2$ is defined using a new type of derivative
$$\pp{}{x_a} : \C Q \rightarrow \C Q \otimes \C Q \qquad x \rightarrow \left( \pp{x}{x_a} \right)^{'} \otimes \left( \pp{x}{x_a} \right)^{''}.$$
We first explain how this derivative acts on paths.  For each occurrence of an arrow $x_a$ in a path, the path can be written as  $x x_a y.$  Split this term into $x \otimes y$ and then sum over all possible positions of the middle arrow.  In Sweedler notation the left part, $x$, is inserted to the first $(\cdot)^{'}$ and the right part, $y,$ is inserted into second $(\cdot)^{''}$.  Using this derivative, the map $\mu_2$ is defined as
\begin{align*}
\mu_2 : & A \otimes_S T_2 \otimes_S A \rightarrow A \otimes_S T_1 \otimes_S A \\
& x \otimes R_a \otimes y \rightarrow \sum_{b \in Q_1} x \left( \pp{R_a}{x_b} \right)^{'} \otimes x_b \otimes \left( \pp{R_a}{x_b} \right)^{''} y.
\end{align*}
Finally, the map $\mu_3$ is defined as
\begin{align*}
\mu_3 : & A \otimes_S T_3 \otimes_S A \rightarrow A \otimes_S T_2 \otimes_S A \\
& x \otimes W_v \otimes y \rightarrow \sum_{b \in Q_1 | t(b) = v} x x_b \otimes R_b \otimes y - \sum_{b \in Q_1 | h(b) = v} x \otimes R_b \otimes x_b y
\end{align*}
It is simple to check that the composition of two successive maps $\mu_{j} \circ \mu_{j+1} = 0$ so we can form the following complex:
\begin{equation}
\label{cyr}
\resizebox{14cm}{!}{$
\begin{CD}
 0 @>>> A \otimes_S T_3 \otimes_S A @>  \mu_3 >> A \otimes_S T_2 \otimes_S A @>  \mu_2 >> A \otimes_S T_1 \otimes_S A  @>  \mu_1 >> A \otimes_S  A @> \mu_0 >> A @>>> 0 
\end{CD}
$}
\end{equation}
Ginzburg's main result is the following theorem:
\begin{thm}[\cite{ginzburgcy}]
An associative algebra $A$ is Calabi-Yau of dimension three if and only if the complex \eqref{cyr} is exact.
\end{thm}
The notion of Calabi-Yau algebras used in this theorem is defined by an analog of Serre duality.
\begin{defn}[\cite{ginzburgcy}]
A homologically smooth algebra $A$ is said to be {\it Calabi-Yau of dimension $d$} if there is an $A-$bimodule quasi-isomorphism
$f:A \rightarrow A^{!}[d]$ such that $f=f^{!}[d]$.
Here $$M \rightarrow M^{!} := RHom_{A-Bimod}(M,A \otimes A).$$
\end{defn}
We will use the projective resolution \eqref{cyr} to compute the Hilbert series of graded superpontetial algebras.   
\begin{defn}
The Hilbert series of a graded superpotential algebra $A = \bigoplus_{r \in \mathbb{N}} A_r$ is the $Q_0 \times Q_0$ matrix $H(A;t)$ with $(v,w)$ entry
$$H_{v,w}(A;t) = \sum_{r = 0}^{\infty} t^r \dim(e_v A_r e_w).$$
\end{defn}
\begin{thm}[Ginzburg/Bocklandt \cite{ginzburgcy,MR2355031}]
Let $A = \C Q/(\partial W)$ be a superpotential algebra with $W$ homogeneous of degree $d.$  Associate to the quiver the adjacency matrix $M_Q(t)$ with $(v,w)$ entry
$$M_{v,w}(Q;t) = \sum_{a \in \text{arr} (v \rightarrow w)} t^{\deg(a)}.$$  The Hilbert series of $A$ equals
$$H(A;t) = \frac{1}{1 - M_Q(t) + t^d M^{T}_Q(t^{-1}) - t^d}$$
 where $1$ represents the identity matrix.
\end{thm}
In the next section we will introduce non-commutative resolutions of local Calabi-Yau singularities.  These form a large family of Calabi-Yau algebras.  We expect that the condition that a gauge theory is superconformal  implies that the corresponding superpotential algebra is Calabi-Yau of dimension three.
\begin{conj}
A superpotential algebra $A = \C Q/(\partial W)$ with an R-charge assignment $R: Q_1 \rightarrow (0,1]$ such that 
\begin{itemize}
\item Each field of $Q_1$ appears in at least two terms of the superpotential,
\item The superpotential $W$ is homogeneous of degree 2,
\item The NSVZ beta functions in equation \eqref{NSVZ} vanish,
\end{itemize}
is a Calabi-Yau algebra of dimension 3.
\end{conj}
For the special case of dimer models, this conjecture has been proven \cite{Mozgovoy:2008fd, broomthesis, davisonBT}.
\section{Non-Commutative Crepant Resolutions}
\label{NCCR}
Bondal and Orlov conjectured that different crepant resolutions  $f_1: Y_1 \rightarrow X$ and $f_2: Y_2 \rightarrow X$ of a local Calabi-Yau singularity $X = \Spec R$ should have equivalent derived categories of coherent sheaves \cite{bondalorlov}.  Van den Bergh gave a new proof of this conjecture in dimension three \cite{MR2057015}, which was motivated by \cite{MR1893007}.  One of his insights was to introduce a non-commutative algebra $A$ as an intermediate object.  
$$D^b(\Coh Y_1) \cong D^b(\Mod -A) \cong D^b(\Coh Y_2).$$
Abstracting the properties of the algebra $A$ led van den Bergh to define non-commutative crepant resolutions.
\begin{defn}[van den Bergh \cite{MR2077594}]
A non-commutative crepant resolution (NCCR) of a Gorenstein ring $R$ is an homologically homogeneous $R$-algebra of the form $A = \End_R(M)$ where $M$ is a reflexive $R$-module.
\end{defn}
In practice, we will work with the slightly weaker, but more accessible, class of non-commutative crepant resolutions given by the next theorem.
\begin{thm}[van den Bergh \cite{MR2077594}]
The algebra
$$A = \End_R(M)$$ 
is a non-commutative crepant resolution of a commutative Gorenstein ring $R$ if
\begin{itemize}
\item $M$ is a reflexive $R$-module,
\item $A$ has finite global dimension,
\item $A$ is a MCM $R$-module.
\end{itemize}
\end{thm}
The second condition is necessary to show that NCCRs of Gorenstein rings of dimension three are Calabi-Yau three algebras.  In this paper, we will focus on NCCRs of the form $A = \End_R(M),$ where $M = \bigoplus_{i=0}^N M_i$ and $M_0 = R.$  Since $\End_R(M_0) \cong R$ we can identify closed loops based at the vertex corresponding to $M_0$ with the elements of $R$, or equivalently the holomorphic functions on the variety $X = \Spec R.$
If we view the algebra $A$ is a quiver gauge theory, each module $M_v$ corresponds to a vertex $v$ of the quiver.  The gauge groups $U(n_v)$ associated to the modules $M_v$ have ranks
$$n_v = N \dim_R M_v$$
where $N$ is the number of D3-branes at the singularity and  $\dim_R M_v$ is the rank of the $R$-module $M_v.$

\section{Volume Minimization}
A-maximization determines the true $R$-symmetry of a superconformal field theory in the IR.  The AdS/CFT dual of this problem is determining the Reeb vector that generates the $U(1)$ isometry of the Sasaki-Einstein geometry.  A geometric dual of $a-$maximization for local toric Calabi-Yau threefolds was found by Martelli, Sparks, and Yau \cite{Martelli:2005tp}.  They showed that the Reeb vector field, and hence the volume of a Sasaki-Einstein metric on the base of a local toric Calabi-Yau cone could be computed by minimizing a function computed from toric data.  Later, they generalized their result to manifolds with only a $(\C^{*})^{\ell}$ symmetry \cite{Martelli:2006yb, Gauntlett:2006vf}.  The basic idea is that the asymptotic growth rate of the number of holomorphic functions on the local Calabi-Yau determines the volume of the Sasaki-Einstein horizon manifold \cite{Bergman:2001qi}.

The equivariant index
$$C(q,X) = \Tr \left\{ q \; \vert \; \mathcal{H}^0(X) \right\}$$
counts the holomorphic functions on $X$ indexed by their charges $q \in (\C^{*})^{\ell}.$   The trace in the definition is of the induced $(\C^{*})^{\ell}$ action defined on the vector space of holomorphic function on $X.$
Let $\zeta_a, a= 1 \dots s$ form a basis for the Lie algebra of $U(1)^{\ell} \subset (\C^{*})^{\ell}$ so we can expand the Reeb vector in components
$\xi = \sum_{a = 1}^{\ell} b_a \zeta_a,$
where $b_a$ are real parameters.  For toric manifolds, the equivariant index reduces to the character
$$C(q, X_{\sigma}) = \sum_{m \in \mathcal{S}_{\sigma}} q^m$$  which counts points in a polyhedral cone $S_{\sigma}$ associated to the toric variety.
The volume of the horizon manifold $L^{2n-1}$ is found by minimizing
$$\Vol[L^{2n-1}](b_a) =  \frac{2 \pi^n}{(n-1)!} \lim_{s \rightarrow 0} s^n C(q_a = e^{- s b_a}, X_{\sigma} )  $$
over all possible values of the Reeb vector.  For the case of interest, $n = 3$ and the volume is
$$\Vol[L^5](b_a) =  \pi^3 \lim_{s \rightarrow 0} s^3 C(q_a = e^{-s b_a}, X )$$
as a function of the Reeb vector.

\section{Hilbert Series}
\label{sec:Hilbert}
In this section, we will show how the volume of a horizon manifold $L^5$ can be computed directly from the quiver describing the dual superconformal field theory.  As explained in section \ref{NCCR}, given a singular local Calabi-Yau $X = \Spec R,$ a noncommutative crepant resolution describes the gauge theory on a stack of D3-branes placed at the singularity of $X.$  If the noncommutative crepant resolution is of the form $A = \End_R(M_0 \oplus \dots \oplus M_{|Q_0| - 1}$ with $M_0 := R$, then the closed loops based at the vertex corresponding to $M_0$ are in bijection with the elements of the ring $R.$

To count paths weighted by R-charge, we simply modify the adjacency matrix to have $(v,w)$ component
$$M_Q(t)_{vw} = \sum_{e \in \text{Arrows}(v \rightarrow w)} t^{R(e)}$$ where $R(e)$ is a trial R-charge for the edge $e.$
Since the superpotential has degree 2, the Hilbert series is
\begin{equation}
\label{eqn:HS}
H(Q;t) = \frac{1}{1 - M_Q(t) + t^2 M^{T}_Q(t^{-1}) - t^2}.
\end{equation}
The $(v,w)$ entry of the Hilbert series counts the number of distinct paths from vertex $v$ to vertex $w$ weighted by R-charge where paths are counted up to F-term equivalence.  Since the module $M_0$ corresponds to vertex $0$ of the quiver, the Hilbert series of $R = \Hom(M_0,M_0)$ is given by the $(0,0)$ entry of the Hilbert series.

To match the Hilbert series to the equivariant index $C(q,X)$ of Martelli, Sparks, and Yau, we recall the precise form of the correspondence between the Reeb vector $\xi$ and the $R$-symmetry.  In their normalization, the weight $\mu$ of a holomorphic function on $X$ is determined by
$\mathcal{L}_{\xi} f = \mu i f$ where $\mathcal{L}_{\xi}$ is the Lie derivative along the Reeb vector field.
The Reeb vector is normalized by demanding that 
$$\mathcal{L}_{\xi} \Omega^{3,0} = 3 i \Omega^{3,0}$$
where $\Omega^{3,0}$ is the no-where vanishing holomorphic three form defined away from the singularity.
The holomorphic functions on $X$ determine eigenfunctions of the Laplacian on its horizon manifold $L^5.$  By carefully performing the Kaluza-Klein reduction, Martelli, Sparks, and Yau show that the scaling dimension $\Delta(\O)$ of a mesonic operator $\O$ in the gauge theory is precisely
$$\Delta = \mu.$$
The superconformal algebra relates the scaling dimensions of chiral primary operators to their R-charge
$$R(\O) = \frac{2}{3} \Delta(\O).$$
Combining these identifications, the volume of the horizon manifold is
$$\Vol[L^5] =   \left( \frac{2 \pi}{3} \right)^3 \lim_{s \rightarrow 0} s^3 H_{0,0}(Q;e^{-s}).$$
\section{Examples}
\label{sec:examples}
\subsection{$\C^3$}
The simplest five-dimensional Sasaki-Einstein manifold is the round five-sphere.  Its metric cone is simply $\C^3.$  The dual gauge theory is $\mathcal{N} = 4$ SYM, which has three adjoint scalar fields $X,Y,Z,$ and superpotential $W = \Tr \left( XYZ - XYZ \right).$  The quiver consists of a single node with three loops corresponding to the three adjoint scalar fields.  Let $a,b,c$ denote the trial $R$-charges for these fields.  The weighted adjacency matrix has the single entry
$$M_Q(t; a,b,c) =
\begin{pmatrix}
t^a + t^b + t^c.
\end{pmatrix}
$$
The Hilbert series is
$$H(Q;t;a,b,c) = \frac{1}{1 - (t^a + t^b + t^c) - (t^{2-a} + t^{2 - b} + t^{2-c}) + t^2}.$$
Imposing the constraint that all the R-charges must sum to 2, we can eliminate $c = 2 - a - b.$  We expand the Hilbert series in $t = e^{-s}$ as
$$s^3 H_{0,0}(Q;e^{-s}) = \frac{1}{ab(2 - a - b)} \frac{1}{s^3} + \O(s).$$
Minimizing the volume over $a$ and $b$ we find that
$$Vol[S^5] = \pi^3$$
which agrees with our choice of normalization.
\subsection{Conifold}
\begin{figure}[h]
\label{fig:conifoldquiver}
\begin{center}
\includegraphics[trim = 10mm 0mm 10mm 0mm, clip, width=12cm]{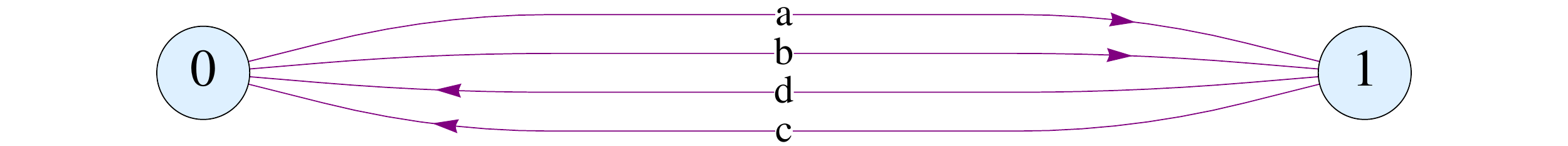}
\caption{Klebanov-Witten quiver for the conifold.}
\end{center}
\end{figure}
The weighted adjacency matrix of the conifold is
$$M_Q(t; a,b,c,d) = 
\begin{pmatrix}
0 & t^a+t^b \\
 t^c+t^d & 0.
\end{pmatrix}
$$
From this we determine the Hilbert series
$$H(Q;t;x,y) = \frac{1-t^2}{\left(1-t^{2-x}\right) \left(1-t^x\right)
   \left(1-t^{2-y}\right) \left(1-t^y\right)}.
$$
We can impose the constraint that the total R-charge is 2 by eliminating $d$ and writing the Hilbert series in terms of $x =  b + c$ and  $y = a+c.$  Expanding the Hilbert series in $t = e^{-s}$ yields
$$ s^3 H_{0,0}(Q; e^{-s}) = \frac{2}{x(2-x)y(2-y)} + \O(s).$$
Minimizing this expression with respect to $x$ and $y,$ we find the volume of the horizon manifold
$$\Vol[T^{1,1}] =  \frac{16 \pi^3}{27}.$$
\section{Perturbative Expansion of the Hilbert Series}
\label{sec:pert}
\subsection{Overview}
In this section, we will prove that the volume formula of Martelli, Sparks, and Yau applied to a quiver arising from a NCCR precisely matches the AdS/CFT prediction from a-maximization.  We will perturbatively expand the Hilbert series $H(Q; t)$ in the variable $t = e^{-s}.$  Our main result is that the expansion takes the form
$$s^3 H_{v,w}(Q; e^{-s})  = s^3 \frac{n_v n_w}{\lambda(s)} + \O(s)$$
where $n_v$ and $n_w$ are the ranks of the gauge groups corresponding to vertices $v$ and $w,$
$$\lambda(s) = \frac{32}{27} a s^3 + \O(s^4),$$
and $a$ is the central charge defined in equation \eqref{eqn:acentral}.  
From this we can compute the volume of the horizon manifold purely in terms of the fields of the quiver gauge theory.
\begin{align*}
\Vol[L^5] & =   \left( \frac{2 \pi}{3} \right)^3 \lim_{s \rightarrow 0} s^3 H_{0,0}(Q;e^{-s}) \\
& =  \left( \frac{2 \pi}{3} \right)^3 \left(  \frac{27}{32} \right) \frac{N^2}{a} \\
& = \frac{\pi^3 N^2}{4a}.
\end{align*}
The volume is precisely as predicted by the AdS/CFT correspondence.

To determine the most singular term in the expansion of $H(Q; e^{-s}),$ we must get control over the eigenvalues of the denominator matrix
$$D_Q(s) \equiv \left( 1 - M_Q(e^{-s}) + e^{-2s} M^{T}_Q(e^{s}) - e^{-2s} \right).$$
By a change of basis, the leading pole in the expansion of $H(Q; e^{-s})$ is governed by the eigenvalue of $D_Q(s)$ with the highest order zero in $s.$ 
Using perturbation theory, we will show there is a unique eigenvalue, $\lambda(s),$ that vanishes as $s^3.$
We begin by Taylor expanding the matrix $D_Q(s),$ the eigenvalue $\lambda(s)$, and its corresponding eigenvector $\ket{\Psi(s)}$ as follows:
\begin{align*}
D_Q(s) & = D_Q^{(0)} + s D_Q^{(1)} + s^2 D_Q^{(2)}  + \dots \\
\ket{\Psi(s)} & = \ket{\Psi^{0}} + s \ket{\Psi^{(1)}} + s^2 \ket{\Psi^{(2)}} + \dots \\
\lambda(s) & = \lambda^{(0)} + s \lambda^{(1)} + s^2 \lambda^{(2)} + \dots
\end{align*}
We first identify the eigenvectors of the leading term $D_Q^{(0)}$ in the expansion.  The $(v,w)$ component of $D_Q(s)$ is
\begin{equation*}
D_{Q}^{vw}(s) =  \left( \sum_{e \in \text{Arr}(v \rightarrow w)} -1 + \sum_{e \in \text{Arr}(w \rightarrow v)} 1 \right) + \O(s).
\end{equation*}
The null vectors of $D_Q^{(0)}$ are spanned by the rank vector $\ket{\phi_0}$ with $v^{th}$ component $n_v$ and baryonic charge vectors $\ket{\phi_J}$ with components $n_v q_v^{J}.$  This follows from our definition of baryonic symmetries as solutions of 
\begin{equation*}
\label{eqn:baryon2}
\sum_{a \in A | h(a) = v} n_{t(a)} q_{t(a)} - \sum_{a \in A | t(a) = v} n_{h(a)} q_{n(a)} = 0.
\end{equation*}
Since $D_Q^{(0)}$ is a real anti-symmetric matrix, we can choose a complete set of orthogonal eigenvectors $\ket{\phi_J}.$  Let $\ket{\phi_0} = \ket{\Psi^{(0)}},$ and label the other
null vectors $\ket{\phi_J}, \; J = 1, \dots, r.$  Label the remaining non-null eigenvectors $ \ket{\phi_J}, \; J = (r+1), \dots, |Q_0| - 1.$
We will show that $\lambda(s) = \frac{32}{27} a s^3 + \O(s^4).$
To accomplish this, we will need the following intermediate results:
\begin{itemize}
\item The rank vector $\ket{\Psi^{(0)}}$ is a null vector of $D_Q^{(0)} + s D_Q^{(1)} .$  We write this as
\begin{equation}
\label{eqn:p1}
\left( D_Q^{(0)} + D_Q^{(1)} s \right) \ket{\Psi^{(0)}} = 0.
\end{equation}
\item 
The order $s^2$ correction to $\lambda(s)$ vanishes.  That is
\begin{equation}
\label{eqn:p2}
\braOket{\Psi^{(0)}}{D_Q^{(2)}}{\Psi^{(0)}} = 0.
\end{equation}
\item The first non-zero correction to $\lambda(s)$ is
\begin{equation}
\label{eqn:p3}
\braOket{\Psi^{(0)}}{D_Q^{(3)}}{\Psi^{(0)}} = \frac{32}{27} a.
\end{equation}
\item The baryonic vectors $\ket{\phi_J}, J = 1 \dots r$ are orthogonal to the rank vector $\ket{\Psi^{(0)}}$ to order $s^3$, that is
\begin{equation}
\label{as:orthog}
\braOket{\phi_J}{D_Q^{(2)}}{\Psi^{(0)}} = 0 \qquad J = 1, \dots r
\end{equation}
\item The matrix governing the mixing of the baryonic symmetries
\begin{equation}
\label{as:negdef}
\braOket{\phi_J}{D_Q^{(1)}}{\phi_K} \qquad J,K = 1, \dots r
\end{equation}
is positive definite.
\end{itemize}
All of these results will follow from general properties of $\mathcal{N} = 1$ superconformal field theories.  We demonstrate properties \eqref{eqn:p1}, \eqref{eqn:p2}, and \eqref{eqn:p3} in section \ref{sec:smallest}.  The remaining two properties, \eqref{as:orthog} and \eqref{as:negdef}, are shown in sections \ref{sec:mixing} and \ref{sec:pos}, respectively.

We expand the eigenvalue equation
\begin{equation}
\label{eqn:eigen}
D_Q(s) \ket{\Psi(s)}  =  \lambda(s)  \ket{\Psi(s)}
\end{equation}
order by order in $s.$
Multiplying through by $\bra{\Psi^{0}}$ on the left and dropping terms of order $\O(s^4)$ we have
\begin{align}
& \bra{\Psi^{0}} \left( D_Q^{(0)} + s D_Q^{(1)} + s^2 D_Q^{(2)} + s^3 D_Q^{(3)} \right) \left(\ket{\Psi^{(0)}} + s \ket{\Psi^{(1)}} + s^2 \ket{\Psi^{(2)}} + s^3 \ket{\Psi^{(3)}} \right) \\
= & s^3  \left( \braOket{ \Psi^{(0)}}{  D_Q^{(3)}}{ \Psi^{(0)}} + \braOket{ \Psi^{(0)}}{  D_Q^{(2)}}{ \Psi^{(1)}}  \right)
\end{align}
where we have used equations \eqref{eqn:p1} and \eqref{eqn:p2}.
For this expression to match the right-hand side of the eigenvalue equation \eqref{eqn:eigen}, $\lambda^{(0)} = \lambda^{(1)} = \lambda^{(2)} = 0,$ and the first non-vanishing correction to $\lambda(s)$ is
\begin{equation}
\label{eqn:shift}
\lambda^{(3)} =  \braOket{ \Psi^{(0)}}{  D_Q^{(3)}}{ \Psi^{(0)}} + \braOket{ \Psi^{(0)}}{  D_Q^{(2)}}{ \Psi^{(1)}}.
\end{equation}
We will show that the first order correction to the eigenvector $\ket{\Psi^{(1)}}$ vanishes and hence
\begin{equation}
\lambda^{(3)} =  \braOket{ \Psi^{(0)}}{  D_Q^{(3)}}{ \Psi^{(0)}} .
\end{equation}

We again expand \eqref{eqn:eigen} perturbatively in $s$ and multiply both sides of the equation by $\bra{\phi_K}.$  Since $\bra{\phi_K}$ was chosen to be a set of mutually orthogonal eigenvectors to the real anti-symmetric matrix, $D_Q^{(0)},$
$\bra{\phi_K} D_Q^{(0)} = - \lambda_K \bra{\phi_K}.$ 
At order $s$ we have the constraint
\begin{align}
\braOket{\phi_K}{D_Q^{(0)}}{\Psi^{(1)}} & = 0 \\
- \lambda_K \braket{\phi_K}{\Psi^{(1)}} & = 0.
\end{align}
where $\lambda_K$ is the corresponding eigenvalue of the eigenvector $\ket{\phi_K}.$
The order $s^2$ term in the expansion is
\begin{equation}
\label{eqn:secorder}
\braOket{\phi_K}{D_Q^{(2)}}{\Psi^{(0)}} + \braOket{\phi_K}{D_Q^{(1)}}{\Psi^{(1)}} + \braOket{\phi_K}{D_Q^{(0)}}{\Psi^{(2)}} = 0.
\end{equation}

We have shown that the first order correction, $\ket{\Psi^{(1)}}$, to $\ket{\Psi(s)}$ must lie in the nullspace of $D_Q^{(0)}.$  The nullspace is spanned by the rank vector $\ket{\Psi^{(0)}}$ and the vectors $\ket{\phi_{J}}, \; J = 1, \dots, r$ associated to the baryonic $U(1)$ symmetries. 

Restricting the basis vectors to the baryonic vectors  $\ket{\phi_{K}}, \; K = 1, \dots, r$  we can further simplify \eqref{eqn:secorder}.  Since the baryonic vectors are in the null space of  of $D_Q^{(0)}, $ equation \eqref{eqn:secorder} reduces to 
\begin{equation}
\braOket{\phi_K}{D_Q^{(2)}}{\Psi^{(0)}} + \braOket{\phi_K}{D_Q^{(1)}}{\Psi^{(1)}} = 0.
\end{equation}

By \eqref{as:orthog}, the first term vanishes.  Furthermore, $ \braOket{\phi_K}{D_Q^{(1)}}{\phi_J}$ is positive definite by \eqref{as:negdef}.  Combined, these two results imply that the leading correction, $\ket{\Psi^{(1)}}$, to the eigenvector $\ket{\Psi(s)}$ must be proportional to $\ket{\Psi^{(0)}}.$  Thus, equation \eqref{eqn:shift} simplifies, and 
\begin{equation}
\lambda^{(3)} =  \braOket{ \Psi^{(0)}}{  D_Q^{(3)}}{ \Psi^{(0)}} 
\end{equation}
as claimed.
All that remains to complete our proof is to show the lemmas given in bullet points.  This will be accomplished in the rest of this section.
\subsection{The Smallest Eigenvalue}
\label{sec:smallest}
Let $t  = e^{-s}$ and perturbatively expand the denominator about $s = 0.$  The $(v,w)$ entry of the denominator matrix is
\begin{equation}
\begin{split}
D_Q^{vw}(s) & =
\left( \sum_{e \in \text{Arr}(v\rightarrow w)} -1 + \sum_{e \in \text{Arr}(w \rightarrow v)} 1 \right)  \\
&+ \left( 2 \delta_{vw} + \sum_{e \in \text{Arr}(v\rightarrow w)} R(e) + \sum_{e \in \text{Arr}(w \rightarrow v)} \left( R(e) - 2 \right) \right) s \\
&+ \left( -2 \delta_{vw} + \sum_{e \in \text{Arr}(v\rightarrow w)} - \frac{R(e)^2}{2} + \sum_{e \in \text{Arr}(w \rightarrow v)}  \left(\frac{R(e)^2}{2} - 2R(e) +2 \right) \right) s^2 \\
&+ \left( \frac{4}{3} \delta_{vw} + \sum_{e \in \text{Arr}(v\rightarrow w)} \frac{R(e)^3}{6} + \sum_{e \in \text{Arr}(w \rightarrow v)} \left(\frac{R(e)^3}{6} - R(e)^2 + 2R(e) - \frac{4}{3} \right) \right) s^3 + \dots
\end{split}
\end{equation}
The function $R(e)$ is the trial R-charge of an edge.  All of the identities we will need to simplify the anomalies with the R-charge will also apply to any trial R-charge $R(e)$ \cite{Intriligator:2003wr}.  The sums $\sum_{e \in \text{Arr}(v \rightarrow w)}$ are over all arrows from vertex $v$ to vertex $w$ in the quiver.  These terms come from expanding $M_Q(e^s).$  The sums over the arrows in the reverse direction arise from expanding $e^{-2s} M_Q^{T}(e^{s})$ and the corresponding summands are the terms in the Taylor expansion of $\exp(s(R(e) - 2)).$  

Vanishing of the triangle anomaly with three gluons \eqref{eqn:rank} implies that the sum of the ranks of the incoming and outgoing arrows at each node vanishes.  This yields the first half of \eqref{eqn:p1},
\begin{equation*}
 D_Q^{(0)} \ket{\Psi^{(0)}} = 0.
\end{equation*}
The second half of \eqref{eqn:p1},
\begin{equation*}
 D_Q^{(1)} \ket{\Psi^{(0)}} = 0
\end{equation*}
follows from \eqref{eqn:rank} and the vanishing of the NSVZ beta function \eqref{NSVZ}.

At order $s^2,$ the rank vector, $\ket{\Psi^{(0)}},$ is not in the null space of $D_Q^{(2)},$ but we can show equation \eqref{eqn:p2}  
\begin{equation*}
\braOket{\Psi^{(0)}}{D_Q^{(2)}}{\Psi^{(0)}} = 0
\end{equation*}
holds by expanding the equation out in components:
\begin{align}
\braOket{\Psi^{(0)}}{D_Q^{(2)}}{\Psi^{(0)}}  & = \left( -2 \sum_{v \in Q_0} n_v^2 + \sum_{e \in \text{Arr}(v \rightarrow w)} - n_v n_w \frac{R(e)^2}{2} + \sum_{e \in \text{Arr}(w \rightarrow v)} n_v n_w  \left(\frac{R(e)^2}{2} - 2R(e) +2 \right) \right)  \notag \\
\label{eqn:wbeta}
 & = \left( -2 \sum_{v \in Q_0} n_v^2 + \sum_{e \in \text{Arr}(w \rightarrow v)} n_v n_w  \left( - 2R(e) +2 \right) \right)  \\
 & = 0. \notag
\end{align}
In going from the first line to the second line, we have used the equality of the number of incoming and outgoing arrows.  The last equality follows from the vanishing of the NSVZ beta functions of the gauge groups.  Finally at order $s^3,$ we show \eqref{eqn:p3}.
\begin{align*}
\braOket{\Psi^{(0)}}{D_Q^{(3)}}{\Psi^{(0)}} & = \frac{1}{3}  \left[ \sum_{v \in Q_0}  4n_v^2 + \sum_{e \in \text{Arr}(w \rightarrow v)} n_v n_w \left((R(e) - 1)^3 + 3(R(e)-1) \right) \right] \\
& = \frac{1}{3}  \left[ \sum_{v \in Q_0}  n_v^2 + \sum_{e \in \text{Arr}(w \rightarrow v)} n_v n_w (R(e) - 1)^3  \right] \\
& = \frac{1}{3} (N_G + \Tr R^3) \\
& = \frac{32}{27} a
\end{align*}
where we have used \eqref{eqn:wbeta} to simplify the second line.  We have found that the smallest eigenvalue is proportional to the a-anomaly.
\subsection{Absence of Mixing}
\label{sec:mixing}
When the quiver gauge theory has baryonic $U(1)$ symmetries, there are additional null vectors, $\ket{\phi_J}, \; J = 1 \dots r,$ of $D_Q^{(0)}.$  In this section we show \eqref{as:orthog}, \begin{equation*}
\braOket{\phi_J}{D_Q^{(2)}}{\Psi^{(0)}} = 0 \qquad J = 1, \dots r
\end{equation*}
which we used to simplify \eqref{eqn:shift}.
Expanding the order $s^2$ term,
\begin{align}
\nonumber
 & -2 \sum_{v \in Q_0} n_v^2 q_v^I - \sum_{e \in \text{Arr}(v \rightarrow w)} n_v n_w q_v^I \frac{R(e)^2}{2} + \sum_{e \in \text{Arr}(w \rightarrow v)} n_v n_w q_v^I \left(\frac{R(e)^2}{2} - 2R(e) +2 \right) \\
 \label{eqn:m1}
 & =   -2 \sum_{v \in Q_0} n_v^2 q_v^I + \sum_{e \in \text{Arr}(w \rightarrow v)} n_v n_w \left((q_v^I - q_w^I) \frac{R(e)^2}{2} - 2 q_v^IR(e) +2q_v^I \right).
\end{align}
To simplify this, we multiply the equation $\hat{\beta}_{1/g_v^2}  = 0$ by $n_v q_v^I$ and sum over the vertices, $v,$ of the quiver.
 \begin{align}
 \nonumber
0 & =   2n_v + \sum_{e \in \text{Arr}(v \rightarrow w)} (R(e)-1) n_w + \sum_{e \in \text{Arr}(w \rightarrow v)} (R(e) - 1) n_w \\
\nonumber
0 & = 2 \sum_{v \in Q_0} n_v^2 q_v^I  + \sum_{e \in \text{Arr}(v \rightarrow w)} n_v n_w q_v^I (R(e)-1)  + \sum_{e \in \text{Arr}(w \rightarrow v)} n_v n_w q_v^I (R(e) - 1)  \\
\label{eqn:m2}
0 & = 2 \sum_{v \in Q_0} n_v^2 q_v^I  + \sum_{e \in \text{Arr}(w \rightarrow v)} n_v n_w q_w^I (R(e)-1) + n_v n_w q_v^I (R(e) - 1)  
\end{align}
Using equation \eqref{eqn:m2}, the quadratic term \eqref{eqn:m1} simplifies to 
\begin{align}
\label{eqn:m3}
& \frac{1}{2} \sum_{e \in \text{Arr}(w \rightarrow v)} n_v n_w (q_v^I - q_w^I)  \left(\frac{R(e)^2}{2} - R(e) + 1 \right).
\end{align}
The constraint $\Tr B^I = 0$ implies $\sum_{e \in \text{Arr}(w \rightarrow v)} n_v n_w (q_v^I - q_w^I) = 0.$  We use this constraint to bring equation \eqref{eqn:m3} to the form
\begin{align*}
& \frac{1}{2} \sum_{e \in \text{Arr}(w \rightarrow v)} n_v n_w (q_v^I - q_w^I) (R(e) - 1)^2  \\
& = \frac{1}{2} \Tr R^2 B^I \\
& = 0
\end{align*}
where we have used the vanishing of the $ \Tr R^2 B^I$ anomaly.
\subsection{Positivity}
\label{sec:pos}
In this section we show that the matrix in \eqref{as:negdef},
\begin{equation*}
\braOket{\phi_J}{D_Q^{(1)}}{\phi_K} \qquad J,K = 1, \dots r
\end{equation*}
is negative definite.  This will complete the proof of our main result.  It is necessary to show this lemma to ensure that $\lambda(s)$ is the only eigenvalue that vanishes as $s^3$.  The new field theory ingredient we will need is that the matrix of trace anomalies, $\Tr R B^{I} B^{J}$ is negative definite.  For a trial R-charge, $\Tr R_t B^{I} B^{J}$ is also negative definite if the trial R-charge is sufficiently close to the true R-charge.  From $\hat{\beta}_{1/g_v^2} = 0$ we can multiply equation \eqref{NSVZ} by $n_v q_v^I q_v^J $ and sum over $v$ to obtain
$$2 \sum_{v \in Q_0} n_v^2 q_v^I q_v^J  +  \sum_{e \in \text{Arr}(v \rightarrow w)} (R(e) - 1) n_v n_w q_v^I q_w^J  +   \sum_{e \in \text{Arr}(w \rightarrow v)} (R(e) - 1) n_v n_w q_v^I q_w^J = 0.$$
From $\Tr B^I = 0$ we can multiply through by $n_v q_v^I q_v^J$ and sum over $v$ to obtain
$$  \sum_{e \in \text{Arr}(v \rightarrow w)}  n_v n_w q_v^I q_w^J = \sum_{e \in \text{Arr}(w \rightarrow v)} n_v n_w q_v^I q_w^J.$$
Using these identities we can simplify
$$\Tr R B^I B^J =  \sum_{e \in \text{Arr}(v \rightarrow w)} n_v n_w(q_v^I - q_w^I)(q_v^J - q_w^J)(R(e) - 1)
+ \sum_{e \in \text{Arr}(w \rightarrow v)} n_v n_w(q_v^I - q_w^I)(q_v^J - q_w^J)(R(e) - 1)$$
to conclude that
$$n_v q_v^I Q_{vw} n_w q_w^J = -\frac{1}{2} \Tr R B^{I} B^{J}.$$
Therefore the matrix
\begin{equation*}
\braOket{\phi_J}{D_Q^{(1)}}{\phi_K} \qquad J,K = 1, \dots r
\end{equation*}
is positive definite since $\Tr R B^{I} B^{J}$ is negative definite.  This completes the proof of our main result.
\section{Conclusion}
We have established the equivalence of a-maximization and volume minimization for $AdS_5 \times L^5$ compactifications where $L^5$ is Sasaki-Einstein whenever the quiver gauge theory is known.  These are the most general supersymmetric compactifications with only self-dual five-form flux.  By restricting to this family of Freund-Rubin compactifications, we have essentially restricted to non-commutative crepant resolutions of the cone $X = C(L^5)$.  However, more general supersymmetric compactifications of the form $AdS_5 \times L^5$ exist.  One famous example is the Pilch-Warner solution \cite{Khavaev:1998fb,Freedman:1999gp,Pilch:2000ej}, which has RR and NS-NS three-form fluxes in addition to the self-dual RR five-form flux.

The most general $\mathcal{N} = 1$ compactification of the form $AdS_5 \times L^5$ with all possible fluxes turned on was considered in \cite{Gauntlett:2005ww}.
These geometries can be systematically studied using generalized complex geometry \cite{Gualtieri:2003dx}.  The volume calculations of Martelli, Sparks, and Yau based on Duistermaat-Heckman localization have been adapted to this setting \cite{Gabella:2009wu}.  These geometries are the natural candidates for duals of general superconformal quiver gauge theories.  Since our computation of the Hilbert series only required the superpotential algebra to be Calabi-Yau of dimension three, it is likely that the equivalence of volume minimization and a-maximization can be extended to this setting.

Generalizing to $AdS_5 \times L^5$ compactifications with all fluxes turned on can be viewed as a non-commutative deformation of the usual AdS/CFT correspondence.  These deformations have been studied in the context of quiver gauge theories, Calabi-Yau algebras, and in supergravity.  Deformations of Calabi-Yau algebras are captured by Hochschild cohomology and correspond to superpotential deformations \cite{Bergman:2006gv, MR2308306}.  A very interesting class of deformations comes from exactly marginal deformations \cite{Leigh:1995ep, Green:2010da, Erkal:2010sh}.  It would be exciting to match exactly marginal deformations of quiver gauge theories to deformations of corresponding generalized complex geometries \cite{Koerber:2006hh}.

Hilbert series play an important role in the computation of the BPS index of multi-trace operators  \cite{Benvenuti:2006qr, Forcella:2008bb}.  Further exploitation of Calabi-Yau algebras \cite{ginzburgcy} may yield new results about the BPS index.  Another closer related index is the $\mathcal{N} = 1$ superconformal index \cite{Romelsberger:2005eg, Kinney:2005ej}.  It is possible that the superconformal index for quiver gauge theories might have a simple expression as well.

Our method of determining the Hilbert series \eqref{eqn:HS}  provides a new way of determining the singularity associated to a quiver gauge theory.  It would be interesting to apply it to gauge theories engineered from branes wrapping obstructed curves \cite{Cachazo:2001gh} \cite{Aspinwall:2010mw}.  We hope that the Hilbert series will help elucidate the structure of  $\mathcal{N} = 1$ superconformal quiver gauge theories.  This would greatly enhance our understanding of the AdS/CFT correspondence.
\section{Acknowledgments}
The author would like to thank Charlie Beil, David Berenstein, Aaron Bergman, Johanna Knapp, David Morrison, and Yuji Tachikawa for helpful discussions and the Institute for the Physics and Mathematics of the Universe for providing an ideal environment for the completion of this work.
This research was supported in part by the National Science Foundation under grants DMS-0606578 and DMS-1007414 and by the World Premier International Research Center Initiative (WPI Initiative), MEXT, Japan.

\bibliography{Amax}

\providecommand{\href}[2]{#2}\begingroup\raggedright\begin{thebibliography}{10}
\addtolength{\parskip}{-1ex}

\bibitem{Morrison:1998cs}
D.~R. Morrison and M.~R. Plesser, ``Non-spherical horizons. {I},'' {\em Adv.
  Theor. Math. Phys.} {\bf 3} (1999)  1--81,
\href{http://arxiv.org/abs/hep-th/9810201}{{\tt hep-th/9810201}}.

\bibitem{Acharya:1998db}
B.~S. Acharya, J.~M. Figueroa-O'Farrill, C.~M. Hull, and B.~J. Spence,
  ``{Branes at conical singularities and holography},'' {\em Adv. Theor. Math.
  Phys.} {\bf 2} (1999)  1249--1286,
\href{http://arxiv.org/abs/hep-th/9808014}{{\tt arXiv:hep-th/9808014}}.

\bibitem{Gubser:1998vd}
S.~S. Gubser, ``Einstein manifolds and conformal field theories,'' {\em Phys.
  Rev.} {\bf D59} (1999)  025006,
\href{http://arxiv.org/abs/hep-th/9807164}{{\tt hep-th/9807164}}.

\bibitem{Henningson:1998gx}
M.~Henningson and K.~Skenderis, ``{The holographic Weyl anomaly},'' {\em JHEP}
  {\bf 07} (1998)  023,
\href{http://arxiv.org/abs/hep-th/9806087}{{\tt arXiv:hep-th/9806087}}.

\bibitem{Intriligator:2003jj}
K.~A. Intriligator and B.~Wecht, ``{The exact superconformal R-symmetry
  maximizes a},'' \href{http://dx.doi.org/10.1016/S0550-3213(03)00459-0}{{\em
  Nucl. Phys.} {\bf B667} (2003)  183--200},
\href{http://arxiv.org/abs/hep-th/0304128}{{\tt arXiv:hep-th/0304128}}.

\bibitem{Martelli:2005tp}
D.~Martelli, J.~Sparks, and S.-T. Yau, ``{The geometric dual of a-maximisation
  for toric Sasaki- Einstein manifolds},''
  \href{http://dx.doi.org/10.1007/s00220-006-0087-0}{{\em Commun. Math. Phys.}
  {\bf 268} (2006)  39--65},
\href{http://arxiv.org/abs/hep-th/0503183}{{\tt arXiv:hep-th/0503183}}.

\bibitem{Martelli:2006yb}
D.~Martelli, J.~Sparks, and S.-T. Yau, ``{Sasaki-Einstein} manifolds and volume
  minimisation,'' \href{http://dx.doi.org/10.1007/s00220-008-0479-4}{{\em
  Commun. Math. Phys.} {\bf 280} (2008)  611--673},
\href{http://arxiv.org/abs/hep-th/0603021}{{\tt arXiv:hep-th/0603021}}.

\bibitem{Bergman:2001qi}
A.~Bergman and C.~P. Herzog, ``{The volume of some non-spherical horizons and
  the AdS/CFT correspondence},'' {\em JHEP} {\bf 01} (2002)  030,
\href{http://arxiv.org/abs/hep-th/0108020}{{\tt arXiv:hep-th/0108020}}.

\bibitem{Butti:2005ps}
A.~Butti and A.~Zaffaroni, ``From toric geometry to quiver gauge theory: The
  equivalence of a-maximization and {Z}-minimization,''
  \href{http://dx.doi.org/10.1002/prop.200510276}{{\em Fortsch. Phys.} {\bf 54}
  (2006)  309--316},
\href{http://arxiv.org/abs/hep-th/0512240}{{\tt arXiv:hep-th/0512240}}.

\bibitem{Lee:2006ru}
S.~Lee and S.-J. Rey, ``{Comments on anomalies and charges of toric-quiver
  duals},'' \href{http://dx.doi.org/10.1088/1126-6708/2006/03/068}{{\em JHEP}
  {\bf 03} (2006)  068},
\href{http://arxiv.org/abs/hep-th/0601223}{{\tt arXiv:hep-th/0601223}}.

\bibitem{Franco:2005rj}
S.~Franco, A.~Hanany, K.~D. Kennaway, D.~Vegh, and B.~Wecht, ``Brane dimers and
  quiver gauge theories,'' {\em JHEP} {\bf 01} (2006)  096,
\href{http://arxiv.org/abs/hep-th/0504110}{{\tt arXiv:hep-th/0504110}}.

\bibitem{tHooft:1993gx}
G.~'t~Hooft, ``{Dimensional reduction in quantum gravity},''
\href{http://arxiv.org/abs/gr-qc/9310026}{{\tt arXiv:gr-qc/9310026}}.

\bibitem{Susskind:1994vu}
L.~Susskind, ``The world as a hologram,'' {\em J. Math. Phys.} {\bf 36} (1995)
  6377--6396,
\href{http://arxiv.org/abs/hep-th/9409089}{{\tt hep-th/9409089}}.

\bibitem{Gauntlett:2006vf}
J.~P. Gauntlett, D.~Martelli, J.~Sparks, and S.-T. Yau, ``Obstructions to the
  existence of {Sasaki-Einstein} metrics,''
  \href{http://dx.doi.org/10.1007/s00220-007-0213-7}{{\em Commun. Math. Phys.}
  {\bf 273} (2007)  803--827},
\href{http://arxiv.org/abs/hep-th/0607080}{{\tt arXiv:hep-th/0607080}}.

\bibitem{Barnes:2005bw}
E.~Barnes, E.~Gorbatov, K.~A. Intriligator, and J.~Wright, ``{Current
  correlators and AdS/CFT geometry},''
  \href{http://dx.doi.org/10.1016/j.nuclphysb.2005.10.013}{{\em Nucl. Phys.}
  {\bf B732} (2006)  89--117},
\href{http://arxiv.org/abs/hep-th/0507146}{{\tt arXiv:hep-th/0507146}}.

\bibitem{Benvenuti:2006xg}
S.~Benvenuti, L.~A. Pando~Zayas, and Y.~Tachikawa, ``{Triangle anomalies from
  Einstein manifolds},'' {\em Adv. Theor. Math. Phys.} {\bf 10} (2006)
  395--432,
\href{http://arxiv.org/abs/hep-th/0601054}{{\tt arXiv:hep-th/0601054}}.

\bibitem{Gubser:1998bc}
S.~S. Gubser, I.~R. Klebanov, and A.~M. Polyakov, ``Gauge theory correlators
  from non-critical string theory,'' {\em Phys. Lett.} {\bf B428} (1998)
  105--114,
\href{http://arxiv.org/abs/hep-th/9802109}{{\tt hep-th/9802109}}.

\bibitem{Witten:1998qj}
E.~Witten, ``Anti-de sitter space and holography,'' {\em Adv. Theor. Math.
  Phys.} {\bf 2} (1998)  253--291,
  \href{http://arxiv.org/abs/hep-th/9802150}{{\tt hep-th/9802150}}.
\url{http://arxiv.org/abs/hep-th/9802150}.

\bibitem{Adler:1969gk}
S.~L. Adler, ``{Axial vector vertex in spinor electrodynamics},''
\href{http://dx.doi.org/10.1103/PhysRev.177.2426}{{\em Phys. Rev.} {\bf 177}
  (1969)  2426--2438}.

\bibitem{Bell:1969ts}
J.~Bell and R.~Jackiw, ``{A PCAC puzzle:} $\pi^0 \rightarrow \gamma \gamma$ in
  the $\sigma$-model,'' \href{http://dx.doi.org/10.1007/BF02823296}{{\em Nuovo
  Cim.} {\bf A60} (1969)  47--61}.

\bibitem{Intriligator:2003wr}
K.~A. Intriligator and B.~Wecht, ``{Baryon charges in 4D superconformal field
  theories and their AdS duals},''
  \href{http://dx.doi.org/10.1007/s00220-003-1023-1}{{\em Commun. Math. Phys.}
  {\bf 245} (2004)  407--424},
\href{http://arxiv.org/abs/hep-th/0305046}{{\tt arXiv:hep-th/0305046}}.

\bibitem{Benvenuti:2004dw}
S.~Benvenuti and A.~Hanany, ``{New results on superconformal quivers},'' {\em
  JHEP} {\bf 04} (2006)  032,
\href{http://arxiv.org/abs/hep-th/0411262}{{\tt arXiv:hep-th/0411262}}.

\bibitem{Novikov:1983ee}
V.~A. Novikov, M.~A. Shifman, A.~I. Vainshtein, and V.~I. Zakharov, ``Instanton
  effects in supersymmetric theories,''
{\em Nucl. Phys.} {\bf B229} (1983)  407.

\bibitem{Douglas:1996sw}
M.~R. Douglas and G.~W. Moore, ``D-branes, quivers, and {ALE} instantons,''
  \href{http://arxiv.org/abs/hep-th/9603167}{{\tt hep-th/9603167}}.
\url{http://arxiv.org/abs/hep-th/9603167}.

\bibitem{Ibanez:1998qp}
L.~E. Ibanez, R.~Rabadan, and A.~M. Uranga, ``{Anomalous U(1)'s in type I and
  type IIB D = 4, N = 1 string vacua},''
  \href{http://dx.doi.org/10.1016/S0550-3213(98)00791-3}{{\em Nucl. Phys.} {\bf
  B542} (1999)  112--138},
\href{http://arxiv.org/abs/hep-th/9808139}{{\tt arXiv:hep-th/9808139}}.

\bibitem{Antoniadis:2002cs}
I.~Antoniadis, E.~Kiritsis, and J.~Rizos, ``{Anomalous U(1)s in type I
  superstring vacua},''
  \href{http://dx.doi.org/10.1016/S0550-3213(02)00458-3}{{\em Nucl. Phys.} {\bf
  B637} (2002)  92--118},
\href{http://arxiv.org/abs/hep-th/0204153}{{\tt arXiv:hep-th/0204153}}.

\bibitem{Jockers:2004yj}
H.~Jockers and J.~Louis, ``{The effective action of D7-branes in N = 1
  Calabi-Yau orientifolds},''
  \href{http://dx.doi.org/10.1016/j.nuclphysb.2004.11.009}{{\em Nucl. Phys.}
  {\bf B705} (2005)  167--211},
\href{http://arxiv.org/abs/hep-th/0409098}{{\tt arXiv:hep-th/0409098}}.

\bibitem{Buican:2006sn}
M.~Buican, D.~Malyshev, D.~R. Morrison, H.~Verlinde, and M.~Wijnholt,
  ``{D-branes at singularities, compactification, and hypercharge},'' {\em
  JHEP} {\bf 01} (2007)  107,
\href{http://arxiv.org/abs/hep-th/0610007}{{\tt arXiv:hep-th/0610007}}.

\bibitem{Martelli:2008cm}
D.~Martelli and J.~Sparks, ``{Symmetry-breaking vacua and baryon condensates in
  AdS/CFT},'' \href{http://dx.doi.org/10.1103/PhysRevD.79.065009}{{\em Phys.
  Rev.} {\bf D79} (2009)  065009},
\href{http://arxiv.org/abs/0804.3999}{{\tt arXiv:0804.3999 [hep-th]}}.

\bibitem{Anselmi:1997am}
D.~Anselmi, D.~Z. Freedman, M.~T. Grisaru, and A.~A. Johansen,
  ``Nonperturbative formulas for central functions of supersymmetric gauge
  theories,'' {\em Nucl. Phys.} {\bf B526} (1998)  543--571,
\href{http://arxiv.org/abs/hep-th/9708042}{{\tt hep-th/9708042}}.

\bibitem{Butti:2005vn}
A.~Butti and A.~Zaffaroni, ``{R-charges from toric diagrams and the equivalence
  of a- maximization and Z-minimization},'' {\em JHEP} {\bf 11} (2005)  019,
\href{http://arxiv.org/abs/hep-th/0506232}{{\tt arXiv:hep-th/0506232}}.

\bibitem{Kutasov:2003ux}
D.~Kutasov, ``{New results on the 'a-theorem' in four dimensional
  supersymmetric field theory},''
\href{http://arxiv.org/abs/hep-th/0312098}{{\tt arXiv:hep-th/0312098}}.

\bibitem{Bertolini:2004xf}
M.~Bertolini, F.~Bigazzi, and A.~L. Cotrone, ``{New checks and subtleties for
  AdS/CFT and a- maximization},''
  \href{http://dx.doi.org/10.1088/1126-6708/2004/12/024}{{\em JHEP} {\bf 12}
  (2004)  024},
\href{http://arxiv.org/abs/hep-th/0411249}{{\tt arXiv:hep-th/0411249}}.

\bibitem{Benvenuti:2004dy}
S.~Benvenuti, S.~Franco, A.~Hanany, D.~Martelli, and J.~Sparks, ``{An infinite
  family of superconformal quiver gauge theories with Sasaki-Einstein duals},''
  {\em JHEP} {\bf 06} (2005)  064,
\href{http://arxiv.org/abs/hep-th/0411264}{{\tt arXiv:hep-th/0411264}}.

\bibitem{Berenstein:2002fi}
D.~Berenstein and M.~R. Douglas, ``{Seiberg duality for quiver gauge
  theories},''
\href{http://arxiv.org/abs/hep-th/0207027}{{\tt arXiv:hep-th/0207027}}.

\bibitem{vdBtalk}
M.~van~den Bergh, ``Introduction to super potentials.'' Oberwolfach talk, 2005.
\newblock \url{http://www.mfo.de/programme/schedule/2005/06/OWR_2005_06.pdf}.

\bibitem{ginzburgcy}
V.~Ginzburg, ``Calabi-yau algebras,''
  \href{http://arxiv.org/abs/math.RA/0612139}{{\tt math.RA/0612139}}.

\bibitem{broomthesis}
N.~Broomhead, {\em Dimer models and Calabi-Yau algebras}.
\newblock PhD thesis, University of Bath, 2009.
\newblock \href{http://arxiv.org/abs/arXiv:0901.4662}{{\tt arXiv:0901.4662}}.

\bibitem{MR1247289}
M.~Kontsevich, ``Formal (non)commutative symplectic geometry,'' in {\em The
  {G}el\cprime fand {M}athematical {S}eminars, 1990--1992}, pp.~173--187.
\newblock Birkh\"auser Boston, Boston, MA, 1993.

\bibitem{MR2355031}
R.~Bocklandt, ``Graded {C}alabi {Y}au algebras of dimension 3,'' {\em J. Pure
  Appl. Algebra} {\bf 212} (2008) no.~1, 14--32,
  \href{http://arxiv.org/abs/math.RA/0603558}{{\tt math.RA/0603558}}.

\bibitem{Mozgovoy:2008fd}
S.~Mozgovoy and M.~Reineke, ``{On the noncommutative Donaldson-Thomas
  invariants arising from brane tilings},''
\href{http://arxiv.org/abs/0809.0117}{{\tt arXiv:0809.0117 [math.AG]}}.

\bibitem{davisonBT}
B.~Davison, ``Consistency conditions for brane tilings,''
  \href{http://arxiv.org/abs/arXiv:0812.4185}{{\tt arXiv:0812.4185}}.

\bibitem{bondalorlov}
A.~I. Bondal and D.~Orlov, ``Semiorthogonal decompositions for algebraic
  varieties,'' \href{http://arxiv.org/abs/alg-geom/9506012}{{\tt
  alg-geom/9506012}}.

\bibitem{MR2057015}
M.~Van~den Bergh, ``Three-dimensional flops and noncommutative rings,'' {\em
  Duke Math. J.} {\bf 122} (2004) no.~3, 423--455,
  \href{http://arxiv.org/abs/math.AG/0207170}{{\tt math.AG/0207170}}.

\bibitem{MR1893007}
T.~Bridgeland, ``Flops and derived categories,'' {\em Invent. Math.} {\bf 147}
  (2002) no.~3, 613--632, \href{http://arxiv.org/abs/math.AG/0009053}{{\tt
  math.AG/0009053}}.

\bibitem{MR2077594}
M.~van~den Bergh, ``Non-commutative crepant resolutions,'' in {\em The legacy
  of {N}iels {H}enrik {A}bel}, pp.~749--770.
\newblock Springer, Berlin, 2004.
\newblock \href{http://arxiv.org/abs/math.RA/0211064}{{\tt math.RA/0211064}}.

\bibitem{Khavaev:1998fb}
A.~Khavaev, K.~Pilch, and N.~P. Warner, ``{New vacua of gauged N = 8
  supergravity in five dimensions},''
  \href{http://dx.doi.org/10.1016/S0370-2693(00)00795-4}{{\em Phys. Lett.} {\bf
  B487} (2000)  14--21},
\href{http://arxiv.org/abs/hep-th/9812035}{{\tt arXiv:hep-th/9812035}}.

\bibitem{Freedman:1999gp}
D.~Z. Freedman, S.~S. Gubser, K.~Pilch, and N.~P. Warner, ``{Renormalization
  group flows from holography supersymmetry and a c-theorem},'' {\em Adv.
  Theor. Math. Phys.} {\bf 3} (1999)  363--417,
\href{http://arxiv.org/abs/hep-th/9904017}{{\tt arXiv:hep-th/9904017}}.

\bibitem{Pilch:2000ej}
K.~Pilch and N.~P. Warner, ``{A new supersymmetric compactification of chiral
  IIB supergravity},''
  \href{http://dx.doi.org/10.1016/S0370-2693(00)00796-6}{{\em Phys. Lett.} {\bf
  B487} (2000)  22--29},
\href{http://arxiv.org/abs/hep-th/0002192}{{\tt arXiv:hep-th/0002192}}.

\bibitem{Gauntlett:2005ww}
J.~P. Gauntlett, D.~Martelli, J.~Sparks, and D.~Waldram, ``{Supersymmetric
  AdS(5) solutions of type IIB supergravity},''
  \href{http://dx.doi.org/10.1088/0264-9381/23/14/009}{{\em Class. Quant.
  Grav.} {\bf 23} (2006)  4693--4718},
\href{http://arxiv.org/abs/hep-th/0510125}{{\tt arXiv:hep-th/0510125}}.

\bibitem{Gualtieri:2003dx}
M.~Gualtieri, ``{Generalized complex geometry},''
\href{http://arxiv.org/abs/math/0401221}{{\tt arXiv:math/0401221}}.

\bibitem{Gabella:2009wu}
M.~Gabella, J.~P. Gauntlett, E.~Palti, J.~Sparks, and D.~Waldram, ``{AdS(5)}
  solutions of type {IIB} supergravity and generalized complex geometry,''
  \href{http://dx.doi.org/10.1007/s00220-010-1083-y}{{\em Commun. Math. Phys.}
  {\bf 299} (2010)  365--408},
\href{http://arxiv.org/abs/0906.4109}{{\tt arXiv:0906.4109 [hep-th]}}.

\bibitem{Bergman:2006gv}
A.~Bergman, ``{Deformations and D-branes},'' {\em Adv. Theor. Math. Phys.} {\bf
  12} (2008)  781--815,
\href{http://arxiv.org/abs/hep-th/0609225}{{\tt arXiv:hep-th/0609225}}.

\bibitem{MR2308306}
R.~Berger and R.~Taillefer, ``Poincar\'e-{B}irkhoff-{W}itt deformations of
  {C}alabi-{Y}au algebras,'' \href{http://dx.doi.org/10.4171/JNCG/6}{{\em J.
  Noncommut. Geom.} {\bf 1} (2007) no.~2, 241--270}.

\bibitem{Leigh:1995ep}
R.~G. Leigh and M.~J. Strassler, ``Exactly marginal operators and duality in
  four-dimensional {N=1} supersymmetric gauge theory,'' {\em Nucl. Phys.} {\bf
  B447} (1995)  95--136,
\href{http://arxiv.org/abs/hep-th/9503121}{{\tt hep-th/9503121}}.

\bibitem{Green:2010da}
D.~Green, Z.~Komargodski, N.~Seiberg, Y.~Tachikawa, and B.~Wecht, ``Exactly
  marginal deformations and global symmetries,''
  \href{http://dx.doi.org/10.1007/JHEP06(2010)106}{{\em JHEP} {\bf 06} (2010)
  106},
\href{http://arxiv.org/abs/1005.3546}{{\tt arXiv:1005.3546 [hep-th]}}.

\bibitem{Erkal:2010sh}
D.~Erkal and D.~Kutasov, ``a-maximization, global symmetries and {RG} flows,''
\href{http://arxiv.org/abs/1007.2176}{{\tt arXiv:1007.2176 [hep-th]}}.

\bibitem{Koerber:2006hh}
P.~Koerber and L.~Martucci, ``{Deformations of calibrated D-branes in flux
  generalized complex manifolds},'' {\em JHEP} {\bf 12} (2006)  062,
\href{http://arxiv.org/abs/hep-th/0610044}{{\tt arXiv:hep-th/0610044}}.

\bibitem{Benvenuti:2006qr}
S.~Benvenuti, B.~Feng, A.~Hanany, and Y.-H. He, ``Counting {BPS} operators in
  gauge theories: Quivers, syzygies and plethystics,''
\href{http://arxiv.org/abs/hep-th/0608050}{{\tt hep-th/0608050}}.

\bibitem{Forcella:2008bb}
D.~Forcella, A.~Hanany, Y.-H. He, and A.~Zaffaroni, ``{The Master Space of N=1
  Gauge Theories},''
  \href{http://dx.doi.org/10.1088/1126-6708/2008/08/012}{{\em JHEP} {\bf 08}
  (2008)  012},
\href{http://arxiv.org/abs/0801.1585}{{\tt arXiv:0801.1585 [hep-th]}}.

\bibitem{Romelsberger:2005eg}
C.~Romelsberger, ``{Counting chiral primaries in N = 1, d=4 superconformal
  field theories},''
  \href{http://dx.doi.org/10.1016/j.nuclphysb.2006.03.037}{{\em Nucl. Phys.}
  {\bf B747} (2006)  329--353},
\href{http://arxiv.org/abs/hep-th/0510060}{{\tt arXiv:hep-th/0510060}}.

\bibitem{Kinney:2005ej}
J.~Kinney, J.~M. Maldacena, S.~Minwalla, and S.~Raju, ``{An index for 4
  dimensional super conformal theories},''
  \href{http://dx.doi.org/10.1007/s00220-007-0258-7}{{\em Commun. Math. Phys.}
  {\bf 275} (2007)  209--254},
\href{http://arxiv.org/abs/hep-th/0510251}{{\tt arXiv:hep-th/0510251}}.

\bibitem{Cachazo:2001gh}
F.~Cachazo, S.~Katz, and C.~Vafa, ``Geometric transitions and {N = 1} quiver
  theories,''
\href{http://arxiv.org/abs/hep-th/0108120}{{\tt arXiv:hep-th/0108120}}.

\bibitem{Aspinwall:2010mw}
P.~S. Aspinwall and D.~R. Morrison, ``Quivers from matrix factorizations,''
\href{http://arxiv.org/abs/1005.1042}{{\tt arXiv:1005.1042 [hep-th]}}.

\end{thebibliography}\endgroup
\end{document}